\documentclass[twocolumn,prl,tightenlines,superscriptaddress]{revtex4-2}
\usepackage{hyperref}
\usepackage{amsmath,amssymb}
\usepackage[utf8]{inputenc} 			% Encodage UTF-8
\usepackage[T1]{fontenc}				% Encodage du fichier de sortie
\usepackage{graphicx}
\usepackage{float}
\usepackage[usenames,dvipsnames]{xcolor}
\usepackage{tikz}
\usepackage{ulem}

\renewcommand{\div}{\nabla\cdot}
\renewcommand{\vec}[1]{\mathbf{#1}}

\newcommand*\colvec[3][]{
    \begin{pmatrix}\ifx\relax#1\relax\else#1\\\fi#2\\#3\end{pmatrix}
}

\begin{document}

\title{Comment on ``The inconvenient truth about flocks'' by Chen {\it et al}}

\author{Hugues Chat\'{e}}
\affiliation{Service de Physique de l'Etat Condens\'e, CEA, CNRS Universit\'e Paris-Saclay, CEA-Saclay, 91191 Gif-sur-Yvette, France}
\affiliation{Computational Science Research Center, Beijing 100094, China}
\affiliation{Sorbonne Universit\'e, CNRS, Laboratoire de Physique Th\'eorique de la Mati\`ere Condens\'ee, 75005 Paris, France}

\author{Alexandre Solon}
\affiliation{Sorbonne Universit\'e, CNRS, Laboratoire de Physique Th\'eorique de la Mati\`ere Condens\'ee, 75005 Paris, France}

\date{\today}
\begin{abstract}
We hope here to provide the community with a convenient account of our viewpoint on the claims made by Chen {\it et al.}
about our results on two-dimensional polar flocks.
\end{abstract}

\maketitle
A recent preprint by Chen and collaborators ~\cite{maitra_inconvenient_2025} revisits once more the problem of 
the scaling properties of fluctuations in two-dimensional polar flocks. 
Here we briefly comment on some of the claims made there,
where it is argued, in particular, that several of the scaling relations
 we obtained in Ref.~\cite{chate_dynamic_2024}, should not hold. 
We discuss the two main points of disagreement, (i) regarding the structure of the
hydrodynamic equations for the Nambu-Goldstone mode associated to
rotational symmetry breaking, (ii) on the pseudo-Galilean invariance
exhibited by the hydrodynamic equation of constant density flocks. In
both cases, we find that the objections raised in  \cite{maitra_inconvenient_2025} are moot so that
our conclusions stand. We stress again that they are in excellent
agreement with all published numerical results.

{\it Dynamics of the Nambu-Goldstone mode.} For a two-dimensional
system, the direction chosen by the spontaneous breaking of rotational
invariance can be parameterized by an angle $\theta(\vec r,t)$. In
Ref.~\cite{chate_dynamic_2024}, we argued that the deterministic part
of the dynamics of $\theta(\vec r,t)$ must obey a continuity equation,
while the noise is non-conserved so that it reads
\begin{equation}
  \label{eq:NG-theta}
  \partial_t \theta(\vec r,t)=-\div \vec J+\eta(\vec r,t)
\end{equation}
where we assumed the noise $\eta(\vec r,t)$ to be Gaussian, white and
delta-correlated in space. Eq.~(\ref{eq:NG-theta}) is a direct
consequence of the fact that, by definition, $\theta(\vec r,t)$ cannot
be enslaved to another direction. For example, the global direction
$\Theta(t)=\frac{1}{V}\int \theta(\vec r,t) d\vec r$ cannot relax
deterministically to another direction and its dynamics must thus be
pure noise $\partial_t\Theta(t)=\xi(t)$ with Gaussian white noise
$\xi$, which leads directly to Eq.~(\ref{eq:NG-theta}).

Ref.~\cite{maitra_inconvenient_2025} first argues that there are known cases where the
symmetry of Eq.~(\ref{eq:NG-theta}) is not obeyed. However, none of
the examples given (the KPZ equation, and the displacement
field in crystals, smectics and discotics) corresponds to the breaking
of rotational symmetry. Moreover, Martin, Parodi and
Pershan~\cite{martin_unified_1972}, which Chen {\it et al} cite to support their
claim, insist on the contrary that, for the breaking of rotational
symmetry, the dynamics of the Goldstone mode take the form of
Eq.~(\ref{eq:NG-theta}) (see the paragraph before Eq.(2.17)
in~\cite{martin_unified_1972}).

{\it Hydrodynamic equations for the Vicsek class.} The presence or
absence of the symmetry Eq.~(\ref{eq:NG-theta}) has important
consequences for the behavior of the resulting hydrodynamic equations
in the Vicsek universality class, for which the Nambu-Goldstone mode
is coupled to a conserved density field. Concretely, the absence of a
continuity equation in
Ref.~\cite{toner_reanalysis_2012,maitra_inconvenient_2025} allows
terms of the form $\theta\partial_x\delta\rho$ and
$\delta\rho\partial_x\theta$ with different coefficients (with here
the mean direction of the flock assumed to be $\theta=0$, along the
$x$-axis, and $\delta\rho(\vec r,t)$ are the density fluctuations). As
we demonstrated in~\cite{chate_dynamic_2024} (Supplementary
information III.C), such terms, under the effect of fluctuations,
generate a mass term $\partial_t \theta=-\alpha\theta+\cdots$ which
destroys the scale free behavior that the theory is supposed to
capture. Against the evidence, Chen {\it et al} consider this to be
impossible because of rotational invariance. This is certainly correct
for equations that indeed have rotation invariance, like their
equations (IV.5-6)~\cite{maitra_inconvenient_2025}. However, strictly
speaking, Eqs.(IV.5-6) are {\it not} a hydrodynamic theory of
fluctuations since they involve non-linearities of arbitrary order. A
hydrodynamic theory is obtained by expanding around the mean direction
of order and keeping only relevant terms, which yields Eq.(IV.17-18)
in~\cite{maitra_inconvenient_2025} truncated at $n=1$ as they discuss,
or Eq.(2.18,2.28) in~\cite{toner_reanalysis_2012}. These equations are
anisotropic, as expected since the mean direction of order breaks
rotation invariance, so that the unphysical mass term is allowed by
symmetry and will indeed be generated by fluctuations. This remark
constitutes a strong argument in favor of the continuity equation
Eq.~(\ref{eq:NG-theta}), which prevents non-gradient terms from being
generated.

A concrete and testable consequence of the continuity Eq.~(\ref{eq:NG-theta}) is
that the noise is not renormalized, which gives the scaling relation
$z=2\chi+1+\zeta$~\footnote{The exponents are defined such that, for a
  flock ordered along the $x$-axis, the system is scale invariant
  under $y\to by$, $x\to b^\zeta x$, $t\to b^z t$,
  $\theta\to b^\chi \theta$, $\delta\rho\to b^\chi \delta\rho$.}. It
is found to be excellently satisfied in numerical simulations of the
Vicsek model~\cite{mahault_quantitative_2019} with error bars less
than $2$\% on that measurement. In addition, Jentsch and Lee, two of
the authors of Ref.~\cite{maitra_inconvenient_2025}, obtained the same
scaling relation using the non-perturbative renormalization
group~\cite{jentsch_new_2024}. Astonishingly, this paper
is not cited in~\cite{maitra_inconvenient_2025}.

{\it Hydrodynamic equation for constant density flocks.} For this
universality class, in which the Nambu-Goldstone mode is the only
hydrodynamic variable, we are satisfied that Chen {\it et al}
obtain exactly the same hydrodynamic equation as our Eq.(8) in Ref.~\cite{chate_dynamic_2024}:
\begin{equation}
  \label{eq:theta-cd}
  \partial_t \theta+\lambda_y\partial_y \theta^2+\lambda_x\partial_x\theta^3 = D_x\partial_x^2\theta+ D_y\partial_y^2\theta + \sqrt{2\Delta}\xi
\end{equation}
with $\xi$ a Gaussian white noise of unit variance. In doing so, they
acknowledged that the $\lambda_x$ term, omitted in the previous
theory~\cite{toner_birth_2012}, is relevant in the RG sense and thus
needs to be included.

\begin{figure}[t]
  \centering
 \includegraphics[width=\columnwidth]{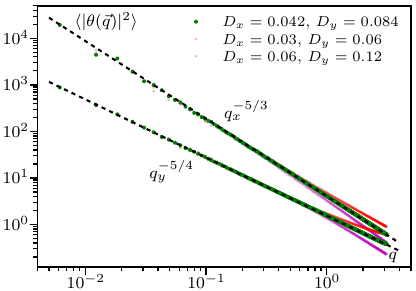}
 \caption{Correlations of fluctuations in $d=2$ constant-density
   flocks Eq.~(\ref{eq:theta-cd}) in the longitudinal $\vec q=(q_x,0)$
   and transverse $\vec q=(0,q_y)$ directions. The dashed lines are
   our theoretical predictions~\cite{chate_dynamic_2024}. The faded
   red and magenta symbols show the short scale effects when going
   away from the optimal value of $D_x$ and $D_y$ (green symbols).
   Parameters: $\lambda_x=1/6$, $\lambda_y=1/2$, $\Delta=0.1$, system
   size $L=1024$ (green) or $L=512$ (red, magenta). Integrated using a
   semi-spectral algorithm with $3/2$ anti-aliasing and Euler
   time-stepping with resolution $dx=1$, $dt=0.005$.}
  \label{fig:corre-cst-density}
\end{figure}

In~\cite{chate_dynamic_2024}, we claimed that, in addition to the
scaling relation $z=2\chi+1+\zeta$ discussed above, two other scaling
relations $\chi+z-1=2\chi-z-\zeta=0$ are given by the
non-renormalization of the $\lambda_x$ and $\lambda_y$ coefficients.
These relations, which Ref.~\cite{maitra_inconvenient_2025} deems
impossible, are extremely well satisfied
numerically. Fig.~\ref{fig:corre-cst-density} shows the correlation
function for $\langle|\theta(\vec q)|^2\rangle$ for wave-vector
$\vec q$, measured in a simulation of Eq.~(\ref{eq:theta-cd}). By
choosing the values of $D_x$ and $D_y$ to minimize the short scale
effects, we observe an outstandingly clean algebraic scaling with the
exponents predicted in Ref.~\cite{chate_dynamic_2024}. Using this new
data, we find that the scaling relations $\chi+z-1=2\chi-z-\zeta=0$
are verified numerically at an accuracy better than $1$\%.

%%%%%%%%%%%%%%%%%%%%%%%%%%%
%%%%%%%%%%%%%%%%%%%%%%%%%%%

We proposed that the non-renormalization of $\lambda_x$ and
$\lambda_y$ may be related to the invariance of
Eq.~(\ref{eq:theta-cd}) under the pseudo-Galilean transformation
\begin{equation}
  \label{eq:galileo}
    \theta'(\vec r',t)=\theta(\vec r-\vec r_c,t)+\theta_0; \;\; \partial_t \vec{r_c}=\theta_0
  \begin{pmatrix}
  6\lambda_x\theta \\ 2\lambda_y
  \end{pmatrix}
\end{equation}
with transformation parameter
$\theta_0$. In~\cite{maitra_inconvenient_2025}, Chen {\it et al}
contested that Eq.~(\ref{eq:galileo}) is a symmetry of
Eq.~(\ref{eq:theta-cd}) because of terms coming from the change of
variable between $\vec r'$ and $\vec r$. We believe that the
discrepancy comes from the fact that Chen {\it et al} considered
arbitrary parameter $\theta_0$. Indeed, such Jacobian terms exist, but
they make contributions at order $\theta_0^2$. They thus disappear for
the infinitesimal $\theta_0$ that we considered
in~\cite{chate_dynamic_2024} (Supplementary information B.1). Let us
stress though, as we did in~\cite{chate_dynamic_2024}, that it is not
clear mathematically that the rather unusual field-dependent
transformation Eq.~(\ref{eq:galileo}) prevents the renormalization of
$\lambda_x$ and $\lambda_y$.

{\it Conclusion.} In this comment, we analyzed several claims of Chen
{\it et al}~\cite{maitra_inconvenient_2025} and found that they are
unjustified. We thus stand by the predictions made
in~\cite{chate_dynamic_2024} which we argued are well supported
analytically and verified numerically to an excellent precision.

\bibliography{refs.bib}

\end{document}